\documentclass[12pt]{article}

\usepackage[centertags]{amsmath}
\usepackage{amsfonts}
\usepackage{amssymb}
\usepackage{amsthm}
\usepackage{amsmath}
\usepackage{graphicx,color}
\usepackage{ulem}
\usepackage{upgreek}

\newcommand{\beq}{\begin{equation}}
\newcommand{\eeq}[1]{\label{#1}\end{equation}}
\newcommand{\bea}{\begin{eqnarray}}
\newcommand{\eea}[1]{\label{#1}\end{eqnarray}}

\newcommand{\ba}{\begin{align}}
\newcommand{\ea}{\end{align}}

\newcommand{\marg}[1]{}

\begin{document}
\setlength{\topmargin}{-1cm} \setlength{\oddsidemargin}{0cm}
\setlength{\evensidemargin}{0cm}
\begin{titlepage}

\noindent
Addentum to Class. Quant. Grav. \textbf{24} (2007), 1683-1686 \cite{Deser:2006sq}

\vspace{2pt}

\noindent
Caltech \# 2021-043

\noindent
Brandeis \# BRX-TH 6695

\vspace{30pt}

\begin{center}
{\Large \bf Higher spin $m\not=0$ excitations on curved backgrounds and cosmological supergravity}

\vspace{25pt}

{\large S. \textsc{Deser}$^{1,2}$ and M. \textsc{Henneaux}$^{3,4}$}

\vspace{25pt}

$^1$Walter Burke Institute for Theoretical Physics Caltech, Pasadena CA 91125, USA; \\
$^2$Physics Department, Brandeis university, Waltham MA 02454, USA; \\
$^3$Physique Th\'eorique et Math\'ematique \& International Solvay Institutes,\\
Universit\'e Libre de Bruxelles, Campus Plaine C.P. 231, B-1050 Bruxelles, Belgium; \\
$^4$Coll\`ege de France, 11 place Marcelin Berthelot, 75005 Paris, France

\end{center}
\vspace{20pt}

\begin{abstract}

Some time ago, we showed that a weak (linear) massless spin 2 wave could only propagate on a Ricci-flat or Ricci-constant background: it must necessarily be a perturbation of General Relativity. This was just the continuation of the higher spin chain: massless $s=3/2$ also required Ricci-flatness (which is the basis of supergravity) while $s>2$ needs Riemann flatness. This note re-analyzes the problem in a perhaps more physical way: by considering  massless and -- only apparently --  massive small excitations and showing how their parameters relate to the cosmological constant. We will thus prove, in a simple physical way, the necessity, as well as the sufficiency, of the known supergravity  results.
\end{abstract}

\end{titlepage}

\newpage

\section{Introduction}\label{sec:intro}

In \cite{Deser:2006sq}, we showed that massless linear $s=2$ excitations, of a very general a priori form, could only propagate on a Ricci --, or equivalently Einstein --, flat or constant space: they had to be effectively perturbations of the general relativity (GR) background. This result fits neatly into a well-known sequence: massive or massless $s< 3/2$ moves in any space, $s=3/2$ only in a Ricci-flat  (if massless) or ``tuned'' constant (if massive) one   (whence supergravity (SUGRA) in the  coupling to gravity), while all higher  than $s=2$  fields require a Riemann-flat background, hence no ``hypergravities''    with $s=5/2$ or higher \cite{Aragone:1979hx}. [Even if $s=5/2$  can live on a fixed, non-dynamical, background such as AdS, that is as unsatisfactory as living in flat space, for the same reason.]

A deeper physics explanation for the above result is that spacetime in GR is not a fixed arena, flat or otherwise, but a dynamical system with its own action. When added to that of spin $2$ $ \sim \int Dh Dh$, the resulting new equations read $G_{\mu \nu}(g) =T_{\mu \nu}(h)$.  Given the linear $\Box h + \cdots =0$  equation, the now obvious result is that the total system is equivalent to  pure GR, but with action $\sim \int R(g+h)$, to quadratic $h$-order. Without this, the spin $2$ field would just be a test field, with negligible $T_{\mu \nu}$, hence of negligible interest. 

A more involved, but similar, story holds for $ s=3/2$ and SUGRA rather than just GR. The combined action of the massless spin $3/2$ field plus GR is consistent because -- and only because --   the ``Bianchi identities'' of the $3/2$ equations, $\sim  (Ricci) D \psi_\mu =0$ hold using $Ricci \sim T_{\mu \nu} (\psi)$ and the rather involved  cubic Fierz identities.  Underlying  this is of course the local supersymmetry of the combined system. Otherwise, the spin $3/2$ field would have to be a dull test-field, with a weak $T_{\mu \nu}$ of no effect as well! [Of course the Einstein Bianchi identities always work, since they merely express local coordinate invariance.]  

In \cite{Deser:2006sq}, we also considered the cosmological case, where identical arguments apply to prove that the background must be  Ricci-constant.  The purpose of this note is to provide physical insight into the derivation of \cite{Deser:2006sq}, by emphasizing how the ``mass'' of the linear $s=2$ excitations is required to connect to the cosmological constant for consistency.

\section{Cosmological excitations}

The basic point is that Cosmological General Relativity (CGR) involves a new, massive, parameter $\Lambda$, so consistent excitations on it must do so as well, simply by now being (apparently) massive. This lesson was learned in SUGRA, where the $s=3/2$ excitation had to become massive to allow the more general SUGRA with cosmological term \cite{Townsend:1977qa}, the upshot being that $s=3/2$ was actually apparently massive as a result of being in a tuned AdS space as explained in \cite{Deser:1977uq}. 

We thus  let the linear $s=2$ wave have an apparent $m^2$  of dimension  $\Lambda$. In the notations and implicitly assumed $(+++-)$ signature of \cite{Deser:2006sq}, this means adding to the Pauli-Fierz action
\begin{equation}
I_2 = \int d^4 x h^{\mu \nu} \theta_{\mu \nu \alpha \beta} h^{\alpha \beta}
\end{equation}
(where $\theta$ is the second-order Hermitian operator yielding the gauge-invariant spin $2$ field equation $\Box h_{\mu \nu} + \cdots = 0$) the term
\begin{equation}
I_C =  -\frac{m^2}{4} \int d^4 x \left(h_{\mu \nu} h^{\mu \nu} - \frac12 h^2 \right) (-g)^{-\frac12}
\end{equation}
where proper background covariance is manifest, but where we do not assume a priori any relationship between $m^2$ and $\Lambda$ (contrary to what was done in \cite{Deser:2006sq}).  Note that the field $h_{\mu \nu}$ carries density weight one, hence the factors $(-g)^{-\frac12}$ in the action \cite{Deser:1969wk}.

However, enforcing gauge invariance of the spin $2$ field under $(-g)^{-\frac12} \delta h_{\mu \nu} = D_\mu \xi_\nu + D_\nu \xi_\mu - g_{\mu \nu} D^\alpha \xi_\alpha$ yields,  by a trivial extension of the calculation leading to Eq. (3) of \cite{Deser:2006sq} that the background must be Ricci-constant with $\Lambda= -m^2$.  Indeed, the variation of the cosmological piece $I_C$ of the action adds to the variation of $I_2[h]$
\begin{equation}
 \delta I_2[h] = \int d^4 x \xi^\mu \left[ R_{\mu \sigma} D_\nu h^{\sigma \nu} + \frac12 (D_\alpha R_{\mu \beta} + D_\beta R_{\mu \beta} - D_\mu R_{\alpha \beta})h^{\alpha \beta} \right] (-g)^{-\frac12}
\end{equation}
the term
\begin{equation}
 \delta I_C[h] =   m^2 \int d^4 x \xi_\mu D_\nu h^{\mu \nu}  (-g)^{-\frac12}
\end{equation}
imposing through $\delta I_2[h] + \delta I_C[h] = 0$ the background equation
\begin{equation}
R_{\mu \nu} = - m^2 g_{\mu \nu}
\end{equation}
equivalent to 
\begin{equation}
G_{\mu \nu} +\Lambda g_{\mu \nu}=0
\end{equation}
with
\begin{equation}
\Lambda = - m^2 \label{eq:LamdaEqualM2}
\end{equation}
(from which $D_\alpha R_{\mu \nu} = 0$ follows automatically). 

We thus see that the background must be indeed Ricci-constant with $\Lambda = - m^2$, so that the total system is equivalent to a pure CGR one with $g_{\mu \nu}+h_{\mu \nu}$, to quadratic $h$ order.
Anti-de Sitter space corresponds to $m^2>0$, i.e., real mass, while de Sitter space corresponds to imaginary mass.  

Of course, as shown in [4], the excitation's ``mass'' effectively vanishes as a result of gauge invariance. For $s=2$ alone, either sign of $m^2$ is allowed, because while dS GR is also effectively massless,  it is finite, involving a geometrical horizon. Real $m$ requires AdS just as in the $s=3/2$ case where $m\sim \sqrt{-\Lambda}$   corresponds to $\Lambda <0$, i.e., anti-deSitter (AdS) gravity in our conventions.  Hence the uniqueness of AdS, vs absence of dS-SUGRA, corresponding to SO(3,2 ) vs SO(4,1)  representations in group language, an alternate explanation.  We stress, however, that the derivation of (\ref{eq:LamdaEqualM2}) given here does not require supersymmetry.

\section{Conclusions}
In summary then, we first rederived our original claim in \cite{Deser:2006sq} that massless s=2 excitations on Einstein-flat/constant spaces in GR are consistent if and only if they are perturbations of the latter, $g_{\mu \nu} \rightarrow g_{\mu \nu}+h_{\mu \nu}$
by a physical argument treating those excitations purely dynamically, without assuming any a priori relationship between the cosmological constant and the mass of the spin $2$ excitation. More centrally, we explained in a dynamical sense, how CGR and SUGRA arise from the corresponding (apparently massive) linear $s=2$ and $3/2$ excitations through the derived relation $\Lambda = - m^2$ in the same way as did their massless counterparts, that is, this condition is not merely sufficient  but necessary.

\section*{Acknowledgments}
The work of SD was supported by the U.S.Department of Energy, Office of Science, Office of High Energy Physics under award number de-sc0011632.  The work of MH was partially supported  by FNRS-Belgium (conventions FRFC PDRT.1025.14 and IISN 4.4503.15) and by funds from the Solvay Family.


\newpage

\noindent
\section{Comment added}

\vspace{.1cm}

\begin{center}
\Large{\bf Maxwell as the geodesic to derive GR}

\vspace{.3cm}


\hspace{1mm} {\large S. \textsc{Deser}}
       
 \end{center}
 
 \vspace{.1cm}






\begin{center}
{\bf Abstract}
\end{center}

	I prove on the basis of a few qualitative observed properties of Maxwell-gravity interaction that gravity is necessarily      GR. Furthermore, because of universality of GR, all matter that interacts with it must therefore also be included.


\subsection{Introduction}
The observational ingredients are well-known: The gravitational field bends light, hence not a scalar. It is macroscopic, hence not spin 1/2 or 3/2. Attractive, so not s=1, leaving spin 2 (all higher spin fields are inconsistent). The theoretical ingredient is a new proof \cite{1} that such a field can only propagate on a Ricci-/Einstein- flat, or constant if the excitation is (apparently) massive, background. The range must be infinite because any finite amount makes the spin 2 acausal \cite{acausal}. Let us recapitulate the history: The initial attempt to derive GR was Kraichnan's \cite{2}, but we recently noted that it was unsatisfactory \cite{3}. Gupta's later attempt was pure handwaving; Feynman's also incomplete - he did not sum the infinite series, but assumed covariance which is equivalent to guaranteeing GR \emph{ab initio}. Weinberg \cite{4} did derive the equations in the sense of the interaction representation, but then the nonlinearity is hidden in the interaction term and the equations are not explicitly visible. The first correct explicit derivation was a one-line one \cite{5} relying on the first order (Palatini) form of GR as a cubic theory, followed by a derivation in an arbitrary background \cite{6} and finally one based on tree-level QM \cite{7}.

\subsection{Proof}
We begin with the source of the spin-2 field, namely the conserved matter stress tensor, strictly speaking that of all matter in the universe, just as Maxwell requires ALL charged currents in order to have conserved sources (even though Maxwell is conformal but the rest of the matter is not. What counts here is conservation, not trace). Now formally covariantize the stress tensor  - this is always possible, and has NO physical content: it just introduces generalized coordinates. Its conservation is now replaced by covariant conservation, still for free. But then the linear spin 2 field's equation on the left side must also be covariantized for consistency. At this point we merely invoke the result of \cite{1}: a spin 2 linear field must necessarily evolve on a Ricci-/Einstein- flat background, of which it is -  importantly - a perturbation. (It could be Riemann-flat as well, but in that case would not be ``dynamical enough'', i.e., would have to be a source-free linear field, which is where we began, pre-coupling!)  So the coupled system can no longer be just covariant in flat space with a fictitious metric, but must live in a Riemann one, subject to the Einstein equations: the metric is now dynamical. Had we allowed the excitation a ``mass'' term, the result for the left side would instead be a perturbation of cosmological GR with constant proportional to $m^2$ [1].  So the only consistent alternative to GR is source-free linear $s=2$ in flat space, dull and unobservable, and no gravity.

\subsection{Acknowledgements}
I thank Marc Henneaux, my collaborator on \cite{1}, of which this note is an application. This work was supported by the U.S. Department of Energy, Office of Science, Office of High Energy Physics, Award Number de-sc0011632.

\bibliographystyle{unsrtnat}


\end{document}